US Army Research Laboratory

# Approaches to Enhancing Cyber Resilience: Report of the North Atlantic Treaty Organization (NATO) Workshop IST-153

by Alexander Kott, Benjamin Blakely, Diane Henshel, Gregory Wehner, James Rowell, Nathaniel Evans, Luis Muñoz-González, Nandi Leslie, Donald W French, Donald Woodard, Kerry Krutilla, Amanda Joyce, Igor Linkov, Carmen Mas-Machuca, Janos Sztipanovits, Hugh Harney, Dennis Kergl, Perri Nejib, Edward Yakabovicz, Steven Noel, Tim Dudman, Pierre Trepagnier, Sowdagar Badesha, and Alfred Møller

# NOTICES

## Disclaimers

US Army Research Laboratory

# Approaches to Enhancing Cyber Resilience: Report of the North Atlantic Treaty Organization (NATO) Workshop IST-153

**by Alexander Kott,** *Office of the Director, ARL*

**Nandi Leslie,** *Raytheon, Waltham, MA*

**Benjamin Blakely, Nathaniel Evans, and Amanda Joyce,** *Argonne National Laboratory, Lemont, IL*

**Diane Henshel and Kerry Krutilla,** *Indiana University, Indianapolis, IL*

**Gregory Wehner and James Rowell,** *US Naval Research Laboratory, Washington, DC*

**Luis Muñoz-González,** *Imperial College, London, UK*

**Donald W French and Donald Woodard,** *Attivo Networks, Fremont, CA*

**Igor Linkov,** *US Army Engineer Research and Development Center, Vicksburg, MS*

**Carmen Mas-Machuca,** *Technische Universität München, Munich, Germany*

**Janos Sztipanovits,** *Vanderbilt University, Nashville, TN*

**Hugh Harney,** *Axiom, Inc., Columbia, MD*

**Dennis Kergl,** *Universität der Bundeswehr, Neubiberg, Germany*

**Perri Nejib and Edward Yakabovicz,** *Northrop Grumman, Falls Church, VA*

**Steven Noel,** *The MITRE Corporation, McLean, VA*

**Pierre Trepagnier,** *MIT Lincoln Laboratory, Lexington, MA*

**Sowdagar Badesha and Tim Dudman,** *Riskaware, Bristol, UK*

**Alfred Møller,** *Danish Defence Acquisition and Logistics Organization, Ballerup, Denmark*

| REPORT DOCUMENTATION PAGE | *Form Approved* *OMB No. 0704-0188* |
|---|---|

| **1. REPORT DATE** *(DD-MM-YYYY)* | **2. REPORT TYPE** | **3. DATES COVERED (From - To)** |
|---|---|---|
| April 2018 | Special Report | 23 October 2017–12 March 2018 |

**4. TITLE AND SUBTITLE**

Approaches to Enhancing Cyber Resilience: Report of the North Atlantic Treaty Organization (NATO) Workshop IST-153

**5a. CONTRACT NUMBER**

**5b. GRANT NUMBER**

**5c. PROGRAM ELEMENT NUMBER**

**6. AUTHOR(S)**

Alexander Kott, Benjamin Blakely, Diane Henshel, Gregory Wehner, James Rowell, Nathaniel Evans, Luis Muñoz-González, Nandi Leslie, Donald W French, Donald Woodard, Kerry Krutilla, Amanda Joyce, Igor Linkov, Carmen Mas-Machuca, Janos Sztipanovits, Hugh Harney, Dennis Kergl, Perri Nejib, Edward Yakabovicz, Steven Noel, Tim Dudman, Pierre Trepagnier, Sowdagar Badesha, and Alfred Møller

**5d. PROJECT NUMBER**

**5e. TASK NUMBER**

**5f. WORK UNIT NUMBER**

**7. PERFORMING ORGANIZATION NAME(S) AND ADDRESS(ES)**

US Army Research Laboratory
Army Research Laboratory (ATTN: RDRL-D)
2800 Powder Mill Road, Adelphi, MD 20783-1138

**8. PERFORMING ORGANIZATION REPORT NUMBER**

ARL-SR-0396

**9. SPONSORING/MONITORING AGENCY NAME(S) AND ADDRESS(ES)**

NATO Science and Technology Organisation
Collaboration Support Office (CSO)
BP 25, 92201 Neuilly sur Seine, France

**10. SPONSOR/MONITOR'S ACRONYM(S)**

NATO

**11. SPONSOR/MONITOR'S REPORT NUMBER(S)**

**12. DISTRIBUTION/AVAILABILITY STATEMENT**

Approved for public release; distribution is unlimited.

**13. SUPPLEMENTARY NOTES**

**14. ABSTRACT**

This report summarizes the discussions and findings of the 2017 North Atlantic Treaty Organization (NATO) Workshop, IST-153, on Cyber Resilience, held in Munich, Germany, on 23–25 October 2017, at the University of Bundeswehr. Despite continual progress in managing risks in the cyber domain, anticipation and prevention of all possible attacks and malfunctions are not feasible for the current or future systems comprising the cyber infrastructure. Therefore, interest in cyber resilience (as opposed to merely risk-based approaches) is increasing rapidly, in literature and in practice. Unlike concepts of risk or robustness—which are often and incorrectly conflated with resilience—resiliency refers to the system's ability to recover or regenerate its performance to a sufficient level after an unexpected impact produces a degradation of its performance. The exact relation among resilience, risk, and robustness has not been well articulated technically. The presentations and discussions at the workshop yielded this report. It focuses on the following topics that the participants of the workshop saw as particularly important: 1) fundamental properties of cyber resilience, 2) approaches to measuring and modeling cyber resilience, 3) mission modeling for cyber resilience, 4) systems engineering for cyber resilience, and 5) dynamic defense as a path toward cyber resilience.

**15. SUBJECT TERMS**

cybersecurity, cyber resilience, risk management, cyber metrics, mission modeling, systems engineering, dynamic defense

| **16. SECURITY CLASSIFICATION OF:** | | | **17. LIMITATION OF ABSTRACT** | **18. NUMBER OF PAGES** | **19a. NAME OF RESPONSIBLE PERSON** |
|---|---|---|---|---|---|
| **a. REPORT** | **b. ABSTRACT** | **c. THIS PAGE** | | | Alexander Kott |
| | | | | | **19b. TELEPHONE NUMBER (Include area code)** |
| Unclassified | Unclassified | Unclassified | UU | 44 | 301-394-1507 |

Standard Form 298 (Rev. 8/98)
Prescribed by ANSI Std. Z39.18

## Contents

## List of Figures

## List of Tables

# Acknowledgments

The co-chairpersons express their gratitude to Mr Julius Zahn for his outstanding help in organizing and executing the workshop, to Mr John B MacLeod for his critical role in organizing the workshop, and to Carol Johnson for organizing and editing this report.

# 1. Introduction

This report summarizes the discussions and findings of the 2017 North Atlantic Treaty Organization (NATO) Workshop, IST-153, on Cyber Resilience. The workshop was held in Munich, Germany, on 23–25 October 2017, at the University of Bundeswehr. This workshop was unclassified and open to NATO nations, Partner for Peace nations, Mediterranean Dialogue, Istanbul Cooperation Initiative nations, and Global Partners. The workshop co-chairpersons were Dr Alexander Kott, US Army Research Laboratory, United States, and Prof Dr Gabrijela Dreo Rodosek, University of Bundeswehr, Germany.

Committee members were Bob Madahar, Defence Science and Technology Laboratory, Cyber and Information Systems Division, United Kingdom; Emin Emrah Özsavaş, Turkish Army Cyber Defence, Turkey; Alfred Møller, Danish Defence Acquisition and Logistics Organization, Denmark; Harald Schmidt, Fraunhofer-Fkie, Germany; Dennis Mccallam, Northrop Grumman, United States; Peeter Lorents, Estonian IT College, Estonia; and Mark Raugas, Pacific Northwest National Laboratory, United States.

In organizing this workshop, the committee stressed that NATO nations—citizens, businesses, and governments—increasingly rely on cyber infrastructure. This puts national security at considerable risk to unforeseen and unknown cyber threats. The high level of interconnectivity found in modern society has opened many avenues for cyberattacks, including internal and external threats, and vulnerabilities within supply-chain networks. Despite continual progress in managing risks in the cyber domain, it is clear that anticipation and prevention of all possible attacks and malfunctions are not feasible for the current or future systems comprising the cyber infrastructure. Therefore, interest in cyber resilience (as opposed to merely risk-based approaches) is increasing rapidly, in literature and in practice, with many nations expressing it in their cyber strategies.

Resilience is defined in dictionaries as the ability to recover from or easily adjust to misfortune or change. It is characterized by 4 abilities: to plan/prepare, absorb, recover from, and adapt to known and unknown threats. Unlike concepts of risk or robustness—which are often and incorrectly conflated with resilience—resiliency refers to the system's ability to recover or regenerate its performance to a sufficient level after an unexpected impact produces a degradation of its performance.

However, the exact relation among resilience, risk, and robustness has not been well articulated technically. This includes the appropriateness and use of metrics, the correspondence to engineering and architectural approaches, and the role of resilience-by-design in assuring effective recovery and continuity of operations. All these issues remain poorly researched and understood.

For all these reasons, the workshop aimed to explore how the directions of current and future science and technology may impact and define potential breakthroughs in this field. The presentations and discussions at the workshop yielded this report. It focuses on the following topics that the participants of the workshop saw as particularly important:

- fundamental properties of cyber resilience
- approaches to measuring and modeling cyber resilience
- mission modeling for cyber resilience
- systems engineering for cyber resilience
- dynamic defense as a path toward cyber resilience

In addition, the papers presented at the workshop were published as a separate volume "Proceedings of the NATO IST-153/RWS-21 Workshop on Cyber Resilience," edited by Alexander Kott and Gabi Dreo Rodosek, found online at http://ceur-ws.org/Vol-2040/.

## 2. Toward a Taxonomy of Fundamental System Properties of Cyber Resilience

Authors: Benjamin Blakely, Diane Henshel, Gregory Wehner, and James Rowell

Cyber resilience is an increasingly discussed but as yet not well-understood concept. Though progress has been made to distinguish it from cyber risk, an exact description of what makes a system resilient and how that resilience can be improved has yet to come. In this section, we build upon the work of Alexeev et al. (2017) to propose specific connections between well-known concepts of information assurance and the resilience of an information system to attack.

We consider risk to be the probable consequence of threat actions (under the model of risk being the product of threat, vulnerability, and consequence), while resilience results in the minimization of the impact of threat actions and the enablement of recovery. The US Department of Homeland Security (DHS 2008) risk lexicon defines these terms similarly, risk as "potential for an unwanted outcome resulting from an incident, event, or occurrence, as determined by its likelihood and the associated consequences" and resilience as the "ability to resist, absorb, recover from, or successfully adapt to adversity or a change in conditions". In a cyber context, we work from the assumption of compromise, famously put by John

Chambers, CEO of Cisco, as “there are 2 types of companies: those that have been hacked and those who don’t know they have been hacked.”

Resilience has been extensively studied in other contexts, and some of those lessons have started to be applied to critical infrastructure. For example, in an ecological context (e.g., Holling [1973]), resilience is a property of a population that can be examined in terms of the properties of equilibria and oscillation given perturbations to the system. Similar thinking has been applied in economics (e.g., Briguglio et al. [2009]). The construction of buildings incorporates resilience properties to resist disasters (Cimellaro et al. 2016). Perhaps most relevant, this type of thinking has been applied to critical infrastructure protection (e.g., Watson [2015]) in an attempt to develop metrics for assessing the resilience of a system.

To understand why a particular system is more resilient than another, we need to understand what properties of a system make it better able to resist, absorb, and recover in the face of an attack. Here we concern ourselves with intentional, malicious threat actors in the context of information technology systems. That system could come in many forms, however—from a single web server, to an interconnected system of microservices, to an embedded system as found in autonomous vehicles.

Alexeev et al. (2017) analyzed resilience at several scales: micro, meso, and macro. We use these scales to categorize system properties that affect its resilience (Table 1). At the microscale, resilience is affected by individual component properties—whether they be software or hardware—and their interfaces with other components. At the mesoscale, architectural properties of the information system, organizational constraints, and operational requirements are used to define the critical resilience properties. At the macroscale, resilience is inherent in the properties of the mission to be carried out by the system(s) and assurance of its successful completion.

**Table 1 Resilience property domains**

| Scale | Domain |
|---|---|
| Microscale | Hardware and software (individual components)<br>Interfaces |
| Mesoscale | Organizational constraints<br>Operational requirements |
| Macroscale | Assurance of mission |

As we wish to focus on the ways systems can be better engineered versus the organization processes surrounding them, we primarily discuss structural resilience. Alexeev et al. (2017) distinguish between structural and active resilience: structural resilience is the ability of a system to resist, absorb, or recover from an attack due to inherent properties in the way it is constructed. However, it is also worth touching on the topic of active resilience—the ability of a system to respond, adapt, and recover (Alexeev et al. 2017). We expect human intervention to provide most of the active resilience actions, but techniques such as moving target defense (MTD) could, in theory, restructure the system in a way that increases its resistance to existing, and perhaps future, attacks.

We separate metrics—the measurement of the behavior of a system, typically in terms of how well it accomplishes its primary functionality—from fundamental properties. Metrics for resilience have been previously studied (e.g., Trimintzios [2011]), and there must be a tight coupling between fundamental properties and metrics. A physical system has properties that allow for prediction of its behavior under certain stresses. However, one would not confuse the spring constant of a particular spring with the metric of how fast the vehicle in which it is installed can go. While this may be an oversimplification, separating these 2 concepts allows us to focus on the resilience properties of a system irrespective of its specific purpose. Returning to Alexeev et al. (2017), we concur with the requirement that properties must be time-independent, quantifiable, and calculated rather than measured.

Therefore, we propose the following mapping of categories of system properties to resilience objectives. For each, we give several examples of traditional information assurance concepts, or modifications of them, that might achieve the desired objective and may be an expression of a fundamental property of the system. From here, we conduct further research into how these, or other, concepts relate to the underlying properties of the system, how to quantify those properties, and how to use them to predict the response of a system to a given attack. Table 2 details the properties of such systems.

**Table 2 Properties of resilient systems**

| | Resist | Absorb | Recover | Respond |
|---|---|---|---|---|
| Hardening | Structural | . . . | . . . | **Active** |
| Diversity and modularity | Structural, **active** | . . . | . . . | **Active** |
| Plasticity | . . . | Structural | . . . | **Active** |
| Graceful degradation | . . . | Structural, **active** | . . . | . . . |
| Instrumentation | . . . | Structural | Structural | **Active** |
| Agility | . . . | Structural | Structural, **active** | **Active** |

A large part of information security, in the cyber domain, focuses on hardening systems to attack at the intersection of risk and resilience. If a system is properly hardened, the vulnerability component of the risk equation is reduced, and thus the probability of a loss event is equivalently reduced, from a probabilistic standpoint. Likewise, if an outlier event occurs, a properly hardened system will be more resistant to an attacker who has penetrated or compromised it. Typically, this is referred to as "defense in depth" and has been the standard model of information security management for at least a decade. Concepts such as mandatory access control (whereby restrictions on even administrative users can be enforced at a granular level, e.g., Loscocco and Smalley [2001]) or Kerckhoff's principle (whereby we assume an attacker has full knowledge of our system and we must maintain security regardless) start with the assumption that an attacker has already penetrated the system and look to mitigate damage. A related concept in light of the widespread impact of ransomware might be persistent access—the ability of an administrator to maintain control of a system despite the actions of a malicious actor.

Similarly, we can resist the actions of a malicious actor by complicating the task of targeting the system. Diversity in system components and their construction (e.g., operating system, programming language, access channel) requires an attacker to formulate multiple attack strategies or be lucky enough to match their strategy to the component presented to them. MTD strategies are designed to increase system diversity. Modularity in a system allows for the reconfiguration and replacement of components without requiring changes to the interface between them, thus enabling greater diversity. Diversity typically creates a level of overhead for administrators and may not be appropriate for all situations. However, in complex modern systems built upon microservice models, interfaces are largely confined to Hypertext Transfer Protocol (HTTP) calls and the underlying details are easily abstracted, allowing diversity between components, if not within them.

Plasticity and graceful degradation allow the system to bend without breaking. Concepts such as redundant components, excess processing or communication capacity, and avoidance of single points of failure are common ways to allow a system to absorb unexpected events while continuing to deliver at least a minimum acceptable level of service. This ability clearly benefits availability, but its effect on confidentiality and integrity is less certain. Graceful degradation is the concept that a system may be preconfigured with a set of progressively less functional states that represent acceptable tradeoffs between continued functionality and the assurance of security parameters. For example, it may be acceptable to proceed with an encrypted connection despite an invalid certification in certain circumstances, or even to fall through to entirely unencrypted communication, but only if predefined criteria are met. Building these considerations into a system represents worst-case thinking, but may be appropriate for highly sensitive or critical functions.

Finally, structural properties of the system, such as instrumentation, can improve the ability of the system's active properties—its ability to recover fully or resiliently continue operations in a degraded but acceptable state. Previous categories were primarily structural; recovery is largely active but depends upon information, that is, adequate instrumentation and monitoring. From a structural perspective, building appropriate instrumentation into the system and monitoring this instrumentation with appropriate reports and alerts (e.g., security information event management tools or intrusion detection/prevention tools) provide information that enables quicker detection of an attack and easier investigation. Whether by manual or automated means, it is much easier to respond to an attack if the right information is available rather than being in the context of uncertainty. The speed with which a system can be repaired or reconstituted may also be affected by structural parameters. Systems built in an "infrastructure as code" paradigm (i.e., with a high level of automation and configuration management) will be much easier to return to normal operation than systems requiring a great deal of manual effort and trial and error. Having frequent backups of critical configuration and data likewise will allow for recovery to a recent known-good state.

## 3. Cyber Resilience-by-Design

Authors: Perri Nejib, Edward Yakabovicz, and Alfred Moller

During the workshop, a cyber system security engineering concept for cyber resilience was introduced to the working group (Fig. 1). The Start Secure, Stay Secure, and Return Secure concept was mapped to the NATO Plan, Prepare, Absorb, Recover, and Adapt concept for cyber resiliency. This addresses the engineering-driven actions necessary to develop more resilient systems by integrating cybersecurity/systems security engineering (SSE) to that of the well-known systems engineering (SE) process. This concept, shown in Fig. 1, infuses SSE techniques, methods, and practices into typical systems and software engineering system development lifecycle activities, thus becoming part of the core solution/process rather than an isolated and expensive add-on, bolt-on, and separate task/process.

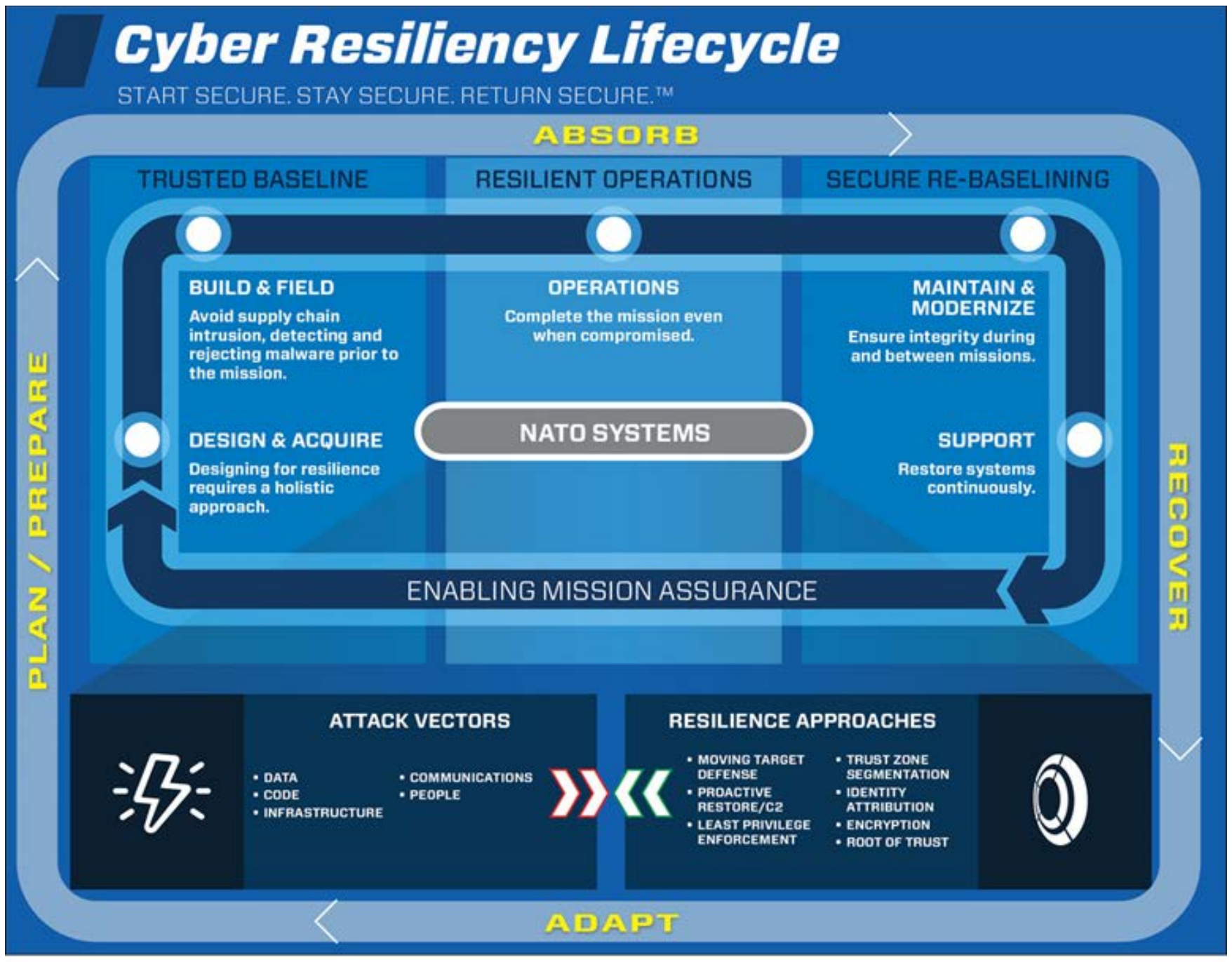

**Fig. 1 NATO cyber-resilience capabilities across the system lifecycle concept—based on Start Secure, Stay Secure, and Return Secure concept**

During the workshop, the group was also introduced to the Cyber Resiliency Engineering Framework developed by MITRE (Bodeau and Graubart 2011), which provides an overview of how to structure cyber-resiliency capabilities by addressing the goals, objectives, and practices in alignment with the "adversary activities" that occur within each capability to reflect the intent and potential actions that the capabilities are intended to protect.

Figure 2 shows the cyber-resilience goals (top) and associated objectives (bottom) from the framework, which align closely with the NATO cyber-resiliency goals. It was apparent throughout the workshop that many of the cyber-resilience concepts presented and introduced were applicable in the NATO domain as well. The following section contains some additional insights that were gained in each of the goal areas during the workshop. It focuses on the architecture perspective as a starting point. Future workshops and papers can expand these data to include other views from engineering, process, and mission assurance perspectives.

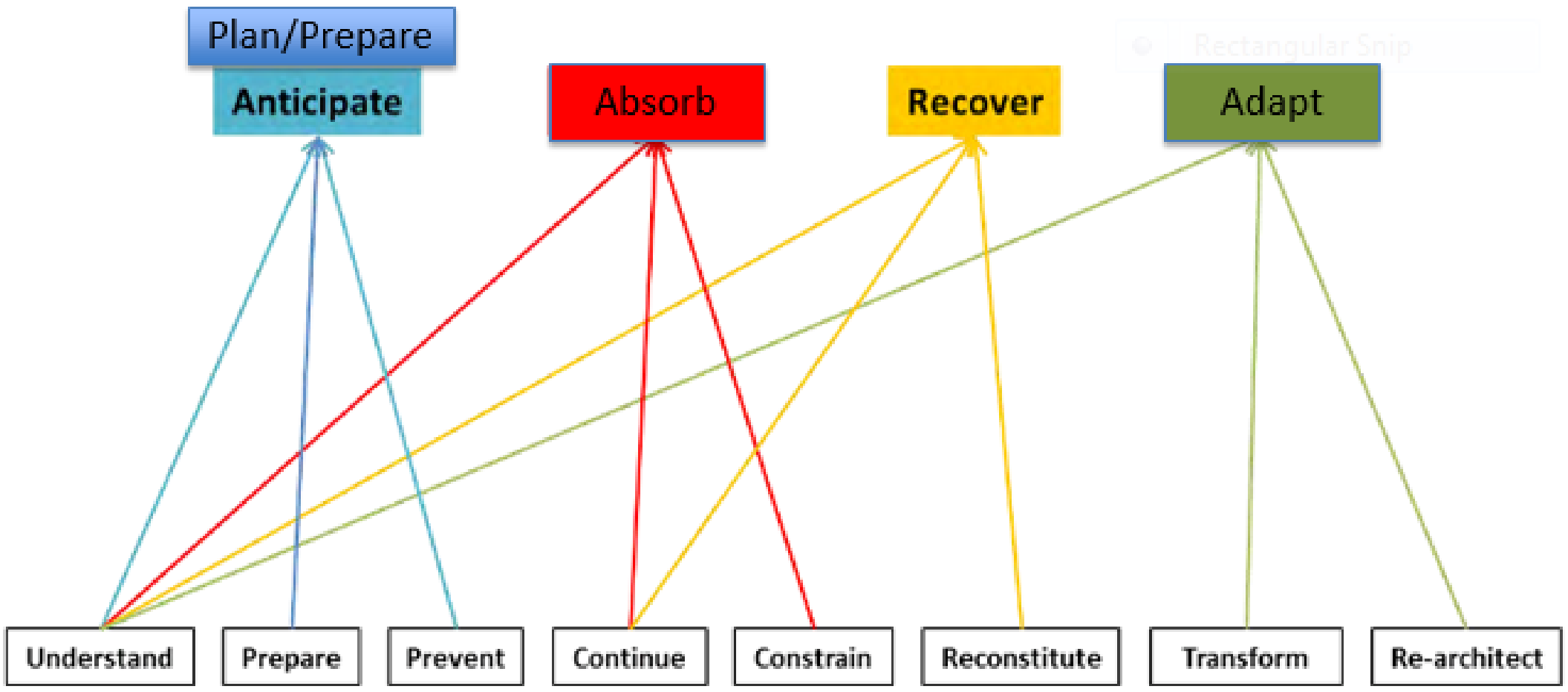

**Fig. 2 Cyber Resiliency Engineering Framework goals and objectives alignment with NATO cyber-resilience goals and objectives (Bodeau and Graubart 2011)**

## 3.1 Goal: Plan/Prepare

For this goal, the architectural approach is reasonably understood and widely used for the design of systems. We have to prepare for cyber resilience by using well-known architecture components, relations, and structures with redundancy, segmentation, diversity, monitoring, coordination, deception, and so on. We also have to be aware of the resilience of the information and communication technology (ICT) platform, because operations are highly dependent on the military platforms.

However, the understanding of relations to operational procedures should be improved. This is prerequisite for having operational people adapt to changes of operations in order to complete a mission even with degraded system capabilities.

## 3.2 Goal: Absorb

The goal of absorbing may seem more difficult from an architectural point of view. Continuation of operations or mission assurance may require unforeseen changeability of the basic system architecture dependent on what is down or degraded by a cyberattack. However, not only the technical systems may require change, but also the operational procedures and way of exchanging information. The consequence may be that alternative communication and processing are necessary together with a change of operational procedures, thus resulting in different mission enterprise architectures, which may not be foreseen in the preparation phase. The result may be that new ways of handling architecture are needed in this dynamic environment.

Constraining the attack is often foreseen during the planning/preparation; however, unexpected ways of attacks may not be foreseen (i.e., some adaptations are needed to isolate faults). How can one use an architectural approach to support this goal of absorbing with limitations of damage, ensuring execution of essential parts of the mission, and understanding of the issues? This may require further investigation and research.

## 3.3 Goal: Recover

The end-state of recovering is normally supported by the architecture. However, transformation from an unexpected state to a recovered state while still maintaining operational continuity is not well understood either and may also require further investigations and research.

## 3.4 Goal: Adapt

The re-architecting phase is often the best understood part of the cyber-resilience architecture process, because it is "easy" to make modifications or reconfigurations based on earlier events or inject emerging technologies for improving the resilience.

As shown in Table 3, we took the framework and again mapped NATO cyber-resilience capabilities to it. We looked closely at the Cyber Resilience Engineering Framework concept and determined that it could be used to describe many of the NATO resiliency attributes, especially taking into account the interdependency among engineering, architecture, and operation/mission. Beyond capabilities, Table 3 also reflects the associated "adversary activities" that occur within each capability that reflect the intent and potential actions from which the capabilities are intended to protect.

**Table 3 Mapping of NATO cyber-resilience capabilities to the Cyber Resilience Engineering Framework**

| Cyber Resilience Capabilities and Goals | System Security Engineering | Architecture | Operations/Mission | Adversary Activities |
|---|---|---|---|---|
| Plan/Prepare (goal: anticipate) Understand-Prepare-Prevent | Enable dynamic reconfiguration and resource re-allocation, using Dynamic Representation and Substantiated Integrity mechanisms that accurately describe the system state<br>Enable dynamic reconstitution, using discovery and Substantiated Integrity mechanisms | Define interfaces with Analytic Monitoring to enable situational awareness of cyber resources and (as feasible) of the surrounding environment and of alternative processing / communications capabilities<br>Define external interfaces to enable situational awareness of the surrounding environment and of alternative processing / communications capabilities. | Involves developing alternative cyber courses of action (CCoAs), and being prepared to respond. Define CCoAs that use externally provided I&W (e.g., DIB tips) Define CCoAs that include I&W thresholds and triggers, as well as damage assessments, using data provided by Analytic Monitoring Define CCoAs that take into consideration mission priorities and constraints on timing of changes | The adversary is preparing the cyber battlefield, seeking to establish a foothold or consolidate a presence in the information infrastructure. The adversary performs reconnaissance, weaponization, and delivery, and attempts exploitation/installation. |
| Absorb (goal: withstand) Understand-Continue-Constrain | Incorporate notification / coordination mechanisms to deconflict actions (e.g., reconfiguration, refresh, resource re-allocation, isolation, failover, reconstitution) by cyber defenders and managers / administrators<br>Design for modularity, so that functional segments can be easily defined<br>Design to separate critical from non-critical data and processing Incorporate thin clients, secure browsers, and diskless nodes to minimize data | Define mappings between the architecture and governance structures, so that those (functional roles and/or architectural components) whose decisions will affect sets of cyber resources are clearly identified<br>Define standards for modularity Provide guidance for defining segments to enable isolation<br>Define standards for trusted, isolated enclaves (criteria or trade-off analyses for when physical separation is needed vs. when virtual enclaves suffice) | The primary focus is thus on maintaining minimal essential capabilities.<br>Define CCoAs that include coordination between cyber defenders and managers or administrators at different tiers or with different spans of control<br>Define CCoAs that isolate mission-essential from non-essential cyber resources | The adversary has established a foothold or consolidated a presence in the information infrastructure, and is using this to subvert the mission (disrupt, deceive, usurp) or compromise future missions (acquire information). The adversary performs command and control and actions to achieve objectives. |
| Recover (goal: recover) Understand-Continue-Reconstitute | Design for agility and interoperability, enabling cyber resources to be repurposed<br>Design for spare capacity and secure failover | Define criteria and trade-offs for realigning resources and functionality Perform trade-off analyses for redundancy, diversity, and costs Provide alternate communications paths for reporting the results of Analytic Monitoring (including indications, warnings, and damage assessments) | When adversary activities are sufficiently contained or defeated, the process of recovering from the attack can begin.<br>Restoration can take the form of backward recovery, rolling back to a known acceptable state.<br>Update CCoAs based on lessons learned from incidents, changes to mission priorities and constraints<br>Define alternate or out-of-band communications / processing paths identified and incorporated | The adversary has demonstrated a presence in or had significant impacts on the information infrastructure, but adversary activities have receded or been curtailed to a tolerable level. The adversary performs maintenance, seeking to ensure future access. |
| Adapt (goal: evolve) Understand-Transform-Re-architect | Design for modularity and agility, so that cyber resources can be relocated, refreshed, and/or replaced | Define standards (criteria and/or trade-offs) for technologies to be replicated, distributed, diversified and/or modularized to facilitate unpredictable location or usage patterns<br>Define standards (criteria and trade-offs) for mission user and cyber defender interfaces that conceal unpredictable behavior that is not relevant to doing their jobs | Environmental changes include changes to the threat environment, the system environment, and the technology environment.<br>Perform realistic exercises that include unpredictable behavior, to evaluate impacts on mission user and cyber defender effectiveness<br>Re-archtecting may include redesigning, re-implementing, or replacing existing cyber resources – particularly with new technologies (innovation), and reconfiguring existing resources to provide new or different capabilities. | The adversary is preparing the cyber battlefield anew, and new adversaries are arising, seeking to establish a foothold or consolidate a presence in the information infrastructure. The adversary seeks intelligence about planned investments in and changes to the information infrastructure (reconnaissance), and may attack the supply chain (weaponization and delivery). |

This section has summarized some of the key findings and discussion points that were a result of the collaboration and exchange during the NATO IST-153 Cyber Resilience Workshop. The table and mapping to the Cyber Resiliency Engineering Framework and the Start Secure, Stay Secure, and Return Secure concept are an excellent starting point toward further development of a NATO cyber resilience-by-design capability.

## 4. Measuring and Modeling Cyber Resilience

Authors: Igor Linkov, Carmen Mas-Machuca, Janos Sztipanovits, Hugh Harney, Dennis Kergl, and Alexander Kott

### 4.1 Challenges of Security in Cyber Domain and Need for Resilience

Analysis, measuring, and modeling of complex systems should include physical, information, cognitive, and social domains (Kott and Abdelzaher 2014). Cyber systems operate within the information domain, as they provide support to mission execution at one end and link to specific operational alternatives at the other end. Generally, cyber systems include sensing, software, and hardware components within systems and networks. These components provide the ability to understand and perform the critical mission functions necessary to sustain operations. Figure 3 outlines the cyber-resilience problem.

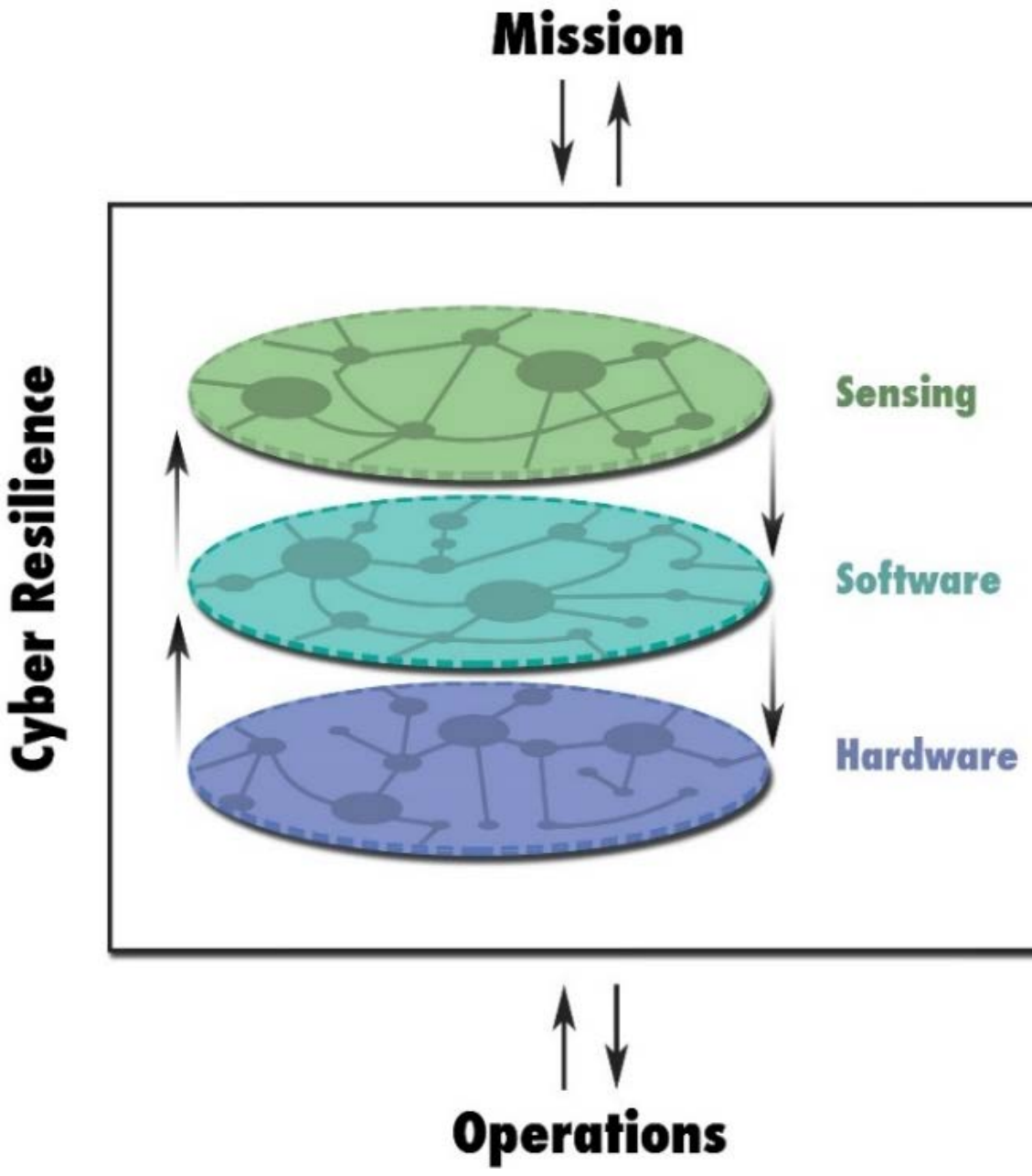

**Fig. 3 Cyber resilience is composed of sensing, hardware, and software networks, which collectively contribute to mission goals and military operations**

The challenges of assessing cybersecurity are related to the interconnected nature of each of the cyber domains and asymmetric threat space. With regard to traditional systems, military personnel can provide protection through the identification of system threats and vulnerabilities as well as system hardening against attacks. This approach is limited for more complex systems in which systems domains are interdependent and interconnected. The hardening of one system component is insufficient to protect against the connected domains, which may result in cascading failures. Given that cyber systems are highly complex, improved methods must account for the interconnected nature of system domains, operations, and goals.

The second challenge of cyber resilience pertains to the adaptive nature of the adversaries themselves. The hardening of system components is highly visible to adversaries who can identify hardened components and subsequently conduct alternative interventions. Thus, while hardening of conventional systems typically results in increased costs of conducting attacks against the network, adversaries of even complex cybersecurity networks can identify barriers and plan counterattacks against other vulnerable components. In the case of the virtual world, adversaries can quickly adapt to overcome system entry barriers at a low cost. Third, the temporal dimension of cyberattacks is short as attacks can be developed in seconds through automated processes and require responses that no longer rely only on human judgment.

To overcome potential network adversaries, resilience ensures that system recovery occurs across each of the physical, information, cognitive, and social domains. Cyber resilience is required to minimize the asymmetric advantage of cyberattacks. To shift from focusing on hardening system components to comprehensive system recovery allows for efficient allocation of resources. It ensures that protection is implemented across all system domains and spatial components to deceive adversaries and ensure high levels of systems protection.

## 4.2 Methods and Tools for Measuring Resilience

Resilience is a new concept in the risk management field. Resilience tools and measures are continually revised to adequately address potential threats to varieties of systems. Figure 4 shows 2 primary resilience approaches that are currently described in the literature: metric based and model based. The core of metric-based approaches entails measures of individual properties of system components using quantitative metrics (e.g., number of antivirus measures, number of system users). While the metrics may or may not be directly related to resilience, metrics should relate to the basic features of resilience, such as the ability of the system to adapt to adversary attacks and systematically recover from disruption. The metrics are traditionally combined either through indices or visualized through dashboards; however, these tools do not necessarily allow for integration of mission goals or tradeoffs within contradictory mission objectives. Thus, a Resilience Matrix (Linkov et al. 2013) provides a more comprehensive approach through the classification of individual metrics based on system domains and temporal evolution of response to threat (absorption, recovery, and adaptation). The Resilience Matrix is capable of integrating multiple metrics by allowing for individual metrics to be weighted through the systematic elicitation of judgment by commanders or analysts.

The second resilience method entails a model-based approach that focuses on representing the real world and defining resilience using models. Modeling requires an understanding of the critical functions of a mission, the critical function thresholds, the temporal patterns of a system, and system memory and adaptation. Process modeling requires a thorough understanding of the physical mechanisms within a system to simulate event impacts and system recovery; process modeling can be difficult to construct and may be information-intensive. Alternatively, statistical approaches require a lot of data on system performance. Bayesian models combine features of process and statistical models. Network models require a presentation of the system as interconnected networks whose structure is dependent on the function of the system. The game-theoretical/agent-based approach focuses on the model performance of the system based on a limited set of rules defined by the modelers. Using these approaches, resilience can be defined. However, the utility of many advanced models is limited due to the data-intensive requirements. Network science is the most promising of the approaches described because the structure of the network can be assessed within the social, cognitive, information, and physical domains of a system.

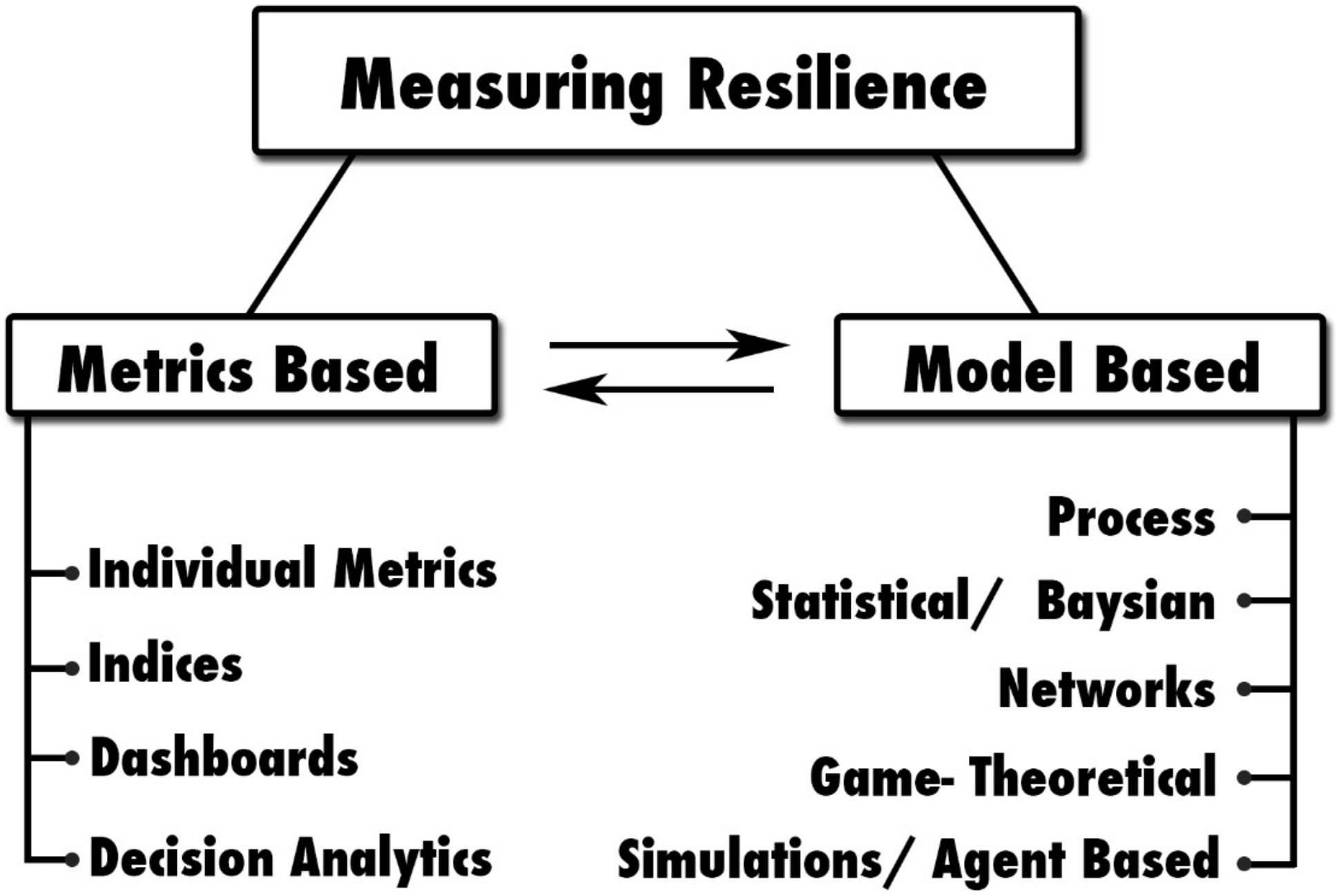

**Fig. 4 Two primary approaches for resilience quantification that are currently described in the literature, including metric- and model-based approaches. Within each of these approaches, multiple tools have been developed to address resilience in systems.**

## 4.3 Example of Measuring Risk and Resilience

Both risk and resilience are important within cybersecurity applications. Risk is useful when the attack vector is known and the vulnerabilities within the system are well defined. Further, risk is useful in scenarios when the consequences of an attack have previously been experienced and are predictable for future threat scenarios. Risk assessment is useful for the identification of specific vulnerabilities affected by system threats. Risk management is useful for the minimization of threat vectors under well-defined threat scenarios. However, risk assessment and management are no longer useful in scenarios in which threats and vulnerabilities are not well defined. Adaptive adversaries will always find a way to find new strategies of attack not currently recognized among the wide range of defense mechanisms. Probability calculations consistent with risk assessment and management may not work in situations of low risk and high uncertainty.

The close relation of risk and resilience has encouraged the research community to take the former as a first step toward the evaluation of resilience. In fact, a system that is able to diminish the risks, reduce the impact of attacks, and improve the response and recovery phase is likely to be more resilient to attacks—though it cannot be guaranteed. The uncertainty justifies why the differences between risk and resilience should be considered, despite their close relation.

Generally, risk models are based on known threats. That is, risk models consider a defined set of failures, attacks, and vulnerabilities that are known for a particular system. Hence, the complete set of adversaries is expected to be well defined (e.g., likelihood to occur, point of impact, impact radius). The resilience models, on the other hand, are designed to be effective given both known and unknown potential threats. Resilience models assess the functions of a particular system to then identify the critical ones. Further, resilience models determine how to restore the performance of the system given an attack in a faster and more efficient way. The restoration of the performance, as shown in Fig. 5, can achieve a level that is lower than, equal to, or even higher than the system performance before the attack. The performance restoration does not have to be done at once, but stepwise. Hence, new parameters measuring the resilience should be defined based on the compromise between the level of performance and the time from the attack. These parameters are depicted in Fig. 6:

- $t_{ip}$ is the time to restore the initial performance,
- $t_{fp}$ is the time to reach the final performance,
- $\Delta p$ is the difference between the final and the initial performance (negative if the system is underperforming), and

- *imp* is the impact of the attack measured by the decrease of performance.

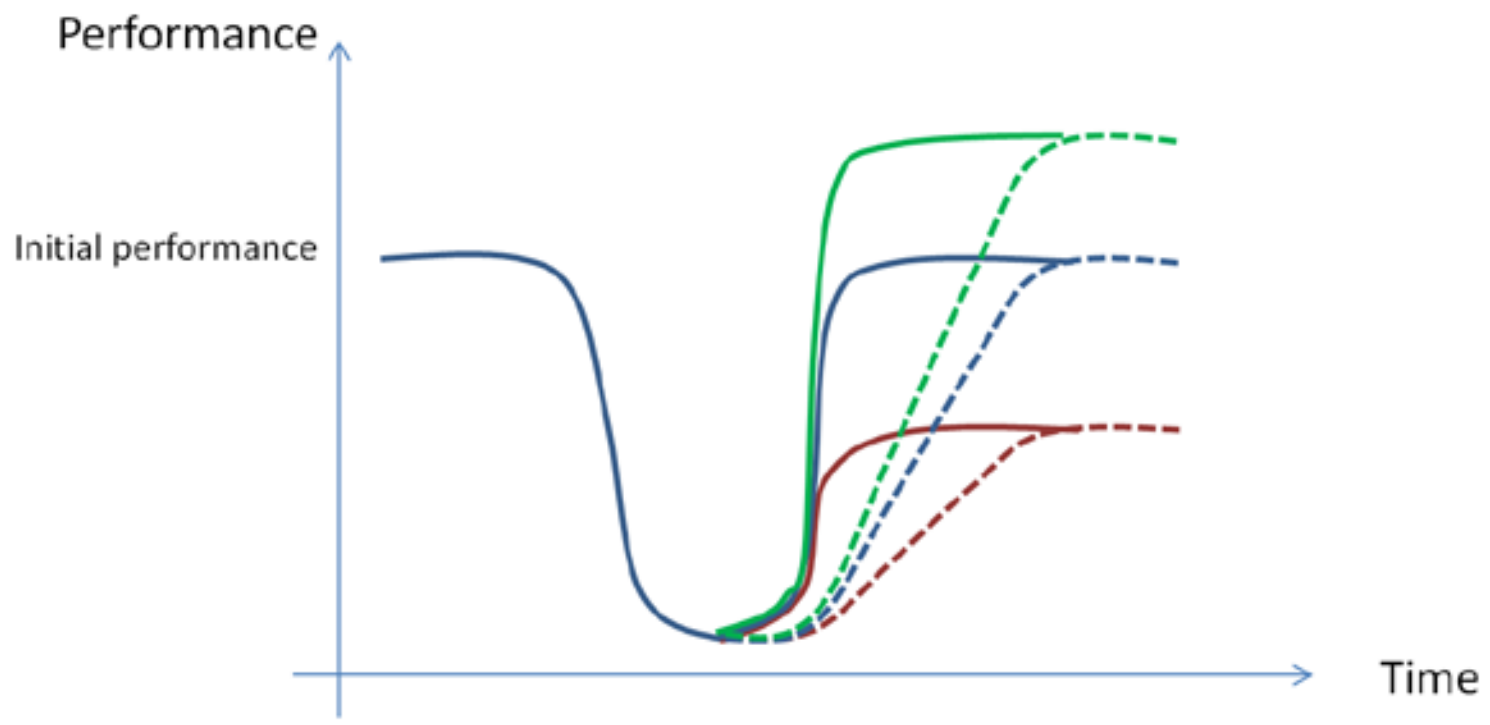

**Fig. 5 Possible resilience profiles: Dashed profiles have worse resilience than the continuous counterparts since it takes longer to restore performance. Comparing performance levels using colors, the green profile offers better performance than the original one, whereas the red profile does not reach the initial performance.**

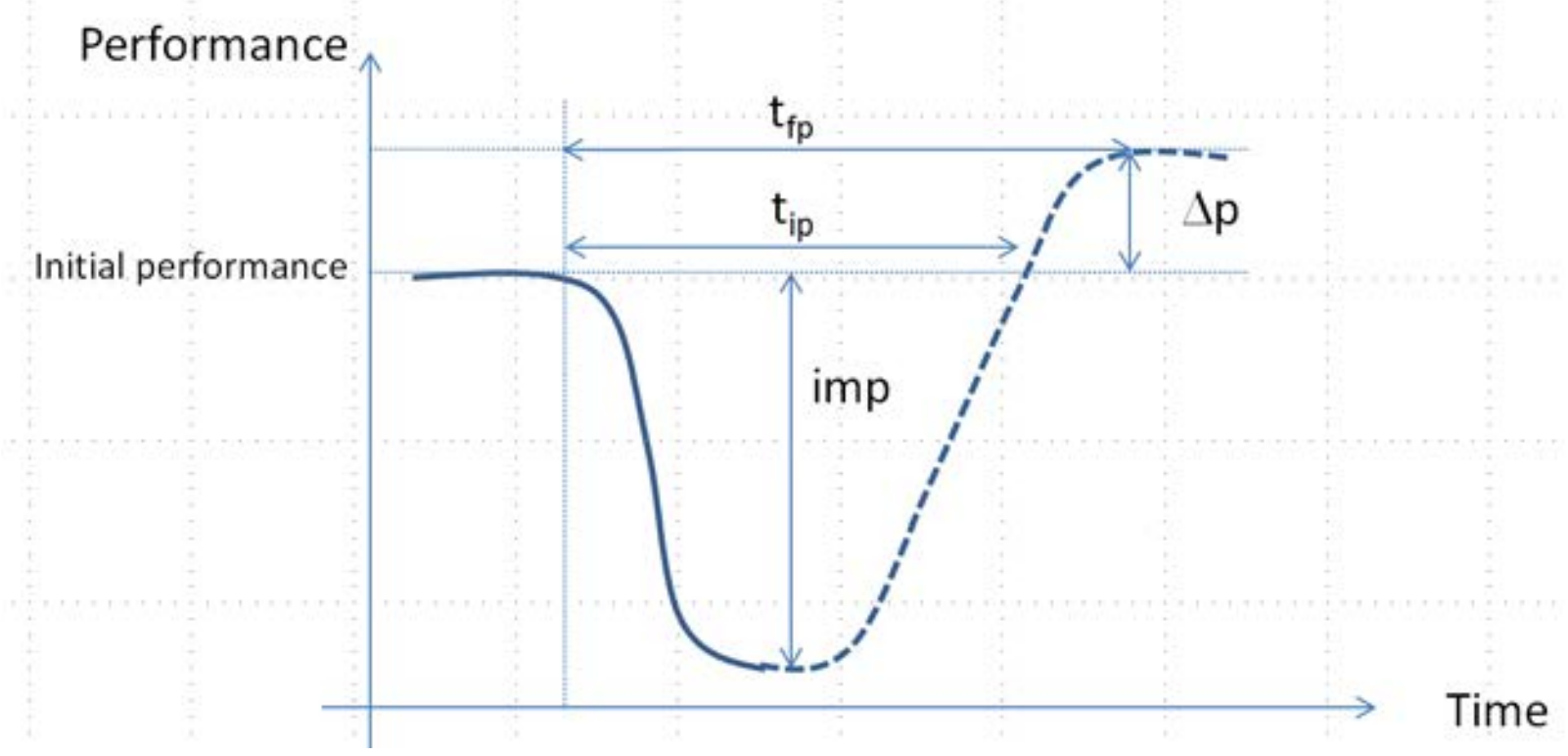

**Fig. 6 Metrics associated to the resilience profile: $t_{ip}$ is the time to restore the initial performance, $t_{fp}$ is the time to reach the final performance, Δp is the difference between the final and the initial performance (negative if the system is underperforming), and *imp* is the impact of the threat measured by the decrease of performance**

To develop an efficient resilience model, a detailed knowledge of the system is required, as well as an understanding of the relationships and dependencies with other systems.

Let us consider a particular example of a communication network in a local area. Assume that, in this case, the communication is provided by different technologies: wireless and wired. The performance could be measured by the number of connections that could survive a particular threat (e.g., loss of power of several communication nodes). Once the attack occurs, several protection and restoration mechanisms are activated so that the maximum and/or the most prioritized connections can be restored. The different mechanisms against an attack with a particular *imp* could be compared in terms of the presented metrics: $t_{ip}$, $t_{fp}$, and Δp.

The concept described previously can also be applied to cyber-physical systems (CPS), where the overall goal is to make the integrated system resilient against cyberattacks, physical attacks, and their combination. To date, resilience of physical systems and cyber systems (including dominantly computation and communication platforms) have been considered as largely disjoint goals in systems engineering that are achieved by different means. In CPSs, where salient functionalities and performance measures emerge from the tight interaction of physical and computation process, this distinction is counterproductive. Resilience needs to be measured in terms of the impact of threats on performance as defined by the context of utilities and risks as they are linked to system vulnerabilities and resources of attackers.

In transportation networks, for example, vehicle throughput can be measured as a performance indicator of the network. Besides the physical topology and characteristics of the road network, the throughput is directly influenced by the traffic control network implemented using cyber means: sensors (inductive loops or imaging), actuators (traffic lights and message boards), and control algorithms (computing nodes and communication networks). The traffic signal schedules are typically designed to maximize throughput and minimize congestion. The risk is manifested in whether attackers exploit cyber vulnerabilities, such as tampering with the schedule of traffic lights in multiple intersections so as to minimize the traffic network utility by maximizing congestion. To avoid detection, attackers may select only valid schedules. In this scenario, the metrics proposed in Fig. 6 can be applied in the following manner.

Assume the initial performance of the traffic network, $T$, that is obtained by optimizing the traffic light schedules. A worst-case attacker that has a certain amount of resources can mount $A$ attacks (including modification of traffic light schedules at most $k$ intersections) simultaneously. The goal of the attacker is to select $A$ such that the throughput $T$ is minimized: $\min_A [T(A)]$. The impact of the attack can be expressed by the metrics: $imp = (T(A)-T)/T$, where $A$ is the worst-case attack given the resources of the attacker.

In this example, we can make the following observations:

1) The dynamics of the attack characterizing the overall resilience are influenced by not only the cyber components (attacker strategy and control network design) but also by the full integrated system dynamics including that of the traffic network.
2) The model of the scenario requires not only physical models of the traffic network but also detailed implementation models of the cyber components and detailed models of the attacks.

3) In dynamic attack scenarios, the full trajectory characterizing attack, impact, and recovery require also modeling the strategies of the attackers and defenders in the framework of the attacker-defender games.

### 4.4 Conclusions

While both risk and resilience methods aim to strengthen a system against potential threats, they have different foundations; therefore, quantification of risk and resilience should be approached differently. Risk assessment is focused on finding a specific asset or system component vulnerable to a known or assumed threat. As the threats and vulnerabilities are known, it is possible to harden the system against these threats. Conversely, resilience models better account for the uncertain threats, vulnerabilities, and consequences. Critical functions of the system are identified, as well as how these functionalities are impacted by unknown or unpredictable threats. That is, resilience models aim to determine whether and how quickly a system's critical functionalities can be restored given an unspecified event in order for the system to recover to a normal, functioning state.

Cybersecurity networks are complex and interconnected networks in which known and unknown threats may impair the integrity of the system as a whole. If the structure of a network's interconnections and critical functionalities is well understood, resilience models help simulate how each state of the network will be impacted by an attack and how cascading effects will influence the resilience of the whole system. While existing risk and resilience modeling approaches are implemented, advanced quantification tools should be further explored to develop a comprehensive understanding of how resilience modeling can help enhance cyber resilience.

## 5. Mission Models for Cyber-Resilient Military Operations

Authors: Steven Noel, Tim Dudman, Pierre Trepagnier, and Sowdagar Badesha

### 5.1 Mission Models and Resilience

In general, cyber resilience is a property of individual systems, system-of-systems, networks, or organizations (Bodeau and Graubart 2017). Understanding cyber risks is a key enabler for achieving appropriate levels of resilience. Because of rich interdependencies among all levels of military activities (operational, tactical, and strategic), cyber risk is not solely determined by individual hosts, vulnerabilities, events, mission functions, and so on. Rather, it is an emergent property that depends on dependencies among entities at all levels of military command and operations. Another dimension of resilience is flexible system components that manage cyber risk by changing configuration and organizational stance, such as modifying boundaries as a reaction to attack, changing roles to meet dependency needs of downstream assets, or forming new communication channels.

Because of the challenges and costs of assessing and improving resilience across a military organization, such activities should be aligned to specific mission requirements. There is a need to identify or discover the various elements that contribute to mission success, and how those elements depend upon each other. This includes 1) high-level mission objectives, 2) the operational tasks that help meet each objective, 3) the system functions that support each task, and 4) the cyber assets that enable each system function. Given the various operational threats and associated risks, such mission dependency models (e.g., graph-based) can guide remediation actions, determine appropriate system redundancies and service diversity, and so on (Noel and Jajodia 2014). Then, as situations involving a cyberattack unfold, mission models can help prioritize alerts, assess elevated mission risk, and understand options for responding to attacks (Musman and Agbolosu-Amison 2014; Noel et al. 2015, 2016). In terms of military doctrine, mission modeling should be considered part of Intelligence Preparation of the Battlefield (HDA 2014).

However, in the current state of practice, developing and continually maintaining mission models remains an expensive process. It might be appropriate to capture some higher-level elements of doctrine through manual efforts (Heinbockel et al. 2016). But because of continual churn at more operational levels, this becomes untenable for lower-level mission elements, especially in tactical environments. As an example, manual methods for producing dependency models from mission threads are expensive and unrepeatable. While there has been some progress in automated methods for mission modeling (Musman 2017; Schulz et al. 2017), there has been relatively little work in this area for tactical environments. Thus, military personnel are continually challenged with understanding how cyberattacks can put missions at risk and impact performance.

## 5.2 Resilience as a Time-Based Problem

Resilience is inherently a time-based problem (Musman et al. 2013; Trepagnier and Schulz 2015). For example, maintaining operational tempo in cyberspace requires synchronizing ongoing analysis with incoming data (e.g., alert streams). At the other end of temporal relevance is the need for aging out data that are no longer relevant to the mission components being protected. Another requirement for cyber resilience is to bring attention to situational changes of relevance to mission assurance, such as deviation from planned versus actual events. For example, such requirements can be addressed through dynamically changing mission-dependency graphs, with re-planning during live missions and activating alternate graph sections in response to events. A system like this could suggest changes that result in more resilient mission graphs (e.g., algorithms that run high impact/low probability analysis).

## 5.3 Time-Based Dependencies

Cyber resilience is quantified by the length of time necessary to recover from a perturbation. In this section, we explore another aspect of time dependence, in which the assets on which a mission depend themselves change over time. Let's explore a toy example taken from the Munich public transportation authority: returning to the Munich airport after the IST-153 Cyber Resilience Workshop. For a specific taxonomy here, we stipulate "Get to the airport from University of the Bundeswehr" to be a sub-mission of the tactical mission "Attending the IST-153 Cyber Resilience Workshop", and that tactical mission to be part of the operational mission "My publications in 2017", which in turn is part of the overarching strategic mission "My successful career in cybersecurity".

As shown in Fig. 7, the tactical sub-mission "Get to the airport from University of the Bundeswehr" contains a baseline course of action: Take Bus 199 from the University gate to the subway station, then take subway U-5 to the Ostbahnhof, and then take S-Bahn 8 from Ostbahnhof to the airport. At the baseline level, mission success depends on the successive availability of the Bus 199, U-5, and S-8 assets. Each of these could be considered as sub-sub-missions of "Get to the airport from University of the Bundeswehr", which in turn is a sub-mission of our tactical mission "Attending the IST-153 Cyber Resilience Workshop".

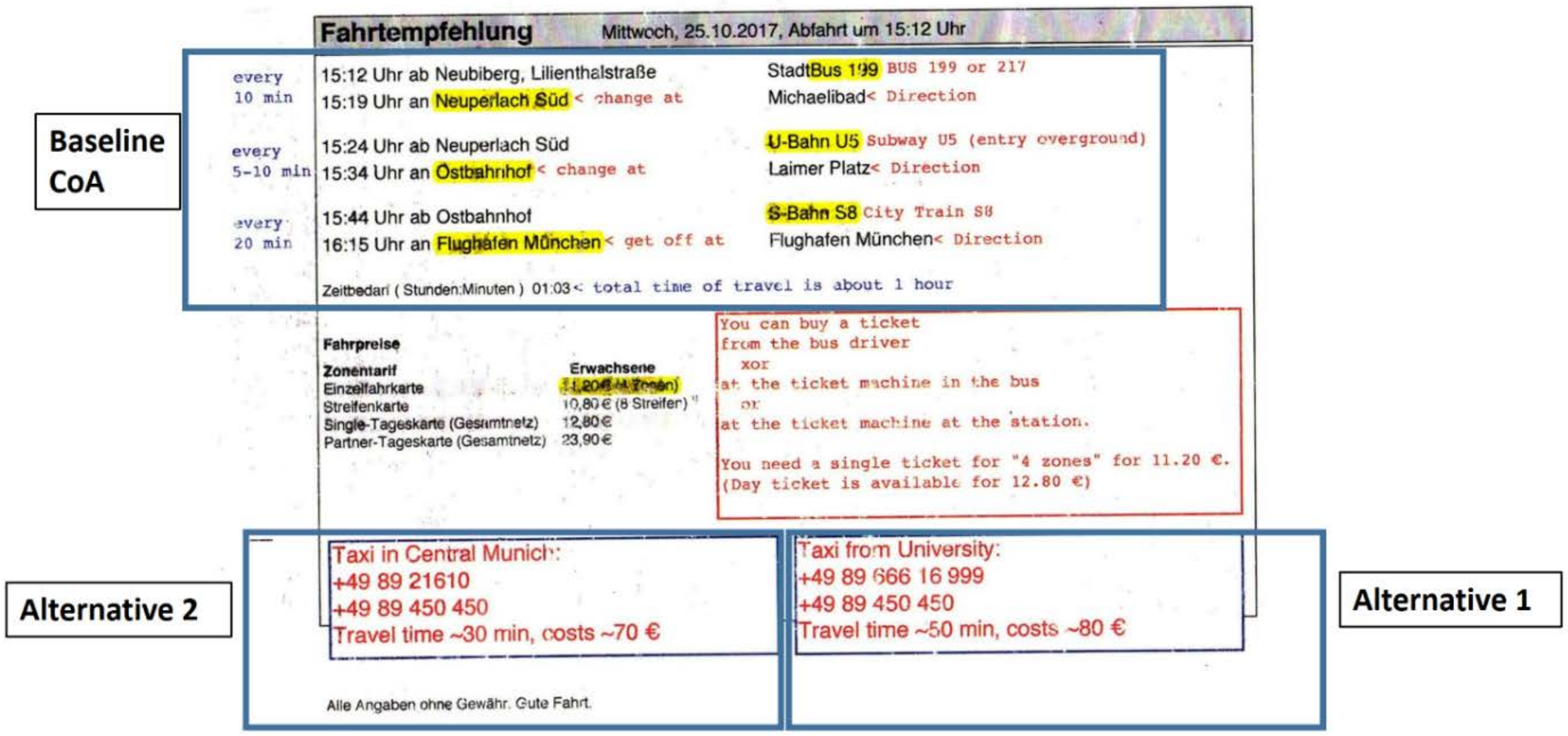

**Fig. 7 “Get to the airport from University of the Bundeswehr” sub-mission**

Note that Alternative 1, “Taxi from the university”, provides resilience in case the bus or U-5 are unavailable, while Alternative 2, “Taxi in central Munich”, would apply in case one has made it to Ostbahnhof, but the S-8 is not available.

The point to emphasize here is the time dependence of the supporting assets regarding their contribution to mission resilience and mission success. Once one makes it to the subway station, the mission dependency on Bus 199 drops out. Similarly, once at Ostbahnhof, U-5 dependency ceases, as does the contribution of Alternative 1 to mission resilience.

## 5.4 Hierarchical Decomposition

The previous discussion suggests a general principle: that, analytically, it makes sense to continue to decompose missions hierarchically until one gets to atomic units where asset dependence is constant over time. (In the sense that the Airport mission dependence on Bus 199 is constant from the moment one steps onto it until one alights at the subway station, and then vanishes.)

We suggest that it also has implications for information filtering. The exact schedules, fares, and phone numbers that the Munich transportation authority has so helpfully provided in Fig. 7 become irrelevant once one has successfully made it to the airport, and need not be passed on up to the next higher level of abstraction. Similarly, the entire process of getting home would be unlikely to be included in a trip report discussing the tactical mission “Attending the IST-153 Cyber Resilience Workshop” and that tactical mission might get only a summary mention in the operational mission “My publications in 2017”, which in turn is subsumed in one’s strategic mission “My successful career in cybersecurity”. Thus, one can avoid the drowning in data issue that would result in retaining the fact at all levels of abstraction that a 4-zone train ticket in Munich costs 11.20 Euros.

## 5.5 Capability Gap

Current methods for assessing mission dependency lack both the granularity and fidelity to apply the hierarchical decomposition approach discussed previously, and more particularly, the discovery of the level of granularity necessary to answer a specific question. Rather than being data-driven, they tend to rely on documentation and the memories and opinions of subject-matter experts. Given constant fixes and patches that occur in cyber, they tend to reflect some combination of the as-designed and as-understood structure, rather than the structure that actually exists at any given time. Developing an accurate method of actually assessing the asset dependencies of a mission, particularly with respect to timescale, is currently a significant unmet need.

## 5.6 Appropriate Modeling Abstractions

When constructing mission models, the model modalities and levels of abstraction need to be matched to the operational use cases. Models need to be sufficiently expressive for answering the required analytic questions and communicating results within an organization. Indeed, as a key part of system resilience, the human element can be included for modeling decisions in response to cyber events. The level of detail needs to be appropriate for the echelon of command, along a spectrum from operational, tactical, and strategic decision making. An important research direction is to develop visual analytics and dashboards appropriate for different command levels.

Mission models based on directed acyclic graphs are built with the Cyber Command System tool (MITRE 2017) and analyzed/visualized via the CyGraph tool (Heinbockel et al. 2016). This captures a hierarchy of dependencies (directed acyclic graph) among mission functions, the information needed for these functions, the services that provide the information, and the hosts on which the services reside. In this way, incorporating mission dependencies supports resilience by prioritizing exploitable paths that lead to mission-critical cyber assets (Musman and Agbolosu-Amison 2014). Cyber resiliency techniques via CyGraph (Noel et al. 2017) are augmented through information on cyber threats, vulnerabilities, policies, and traffic patterns.

An alternative approach is to create mission models that link cyber assets to organizational business processes (Musman et al. 2013; Noel et al. 2015) employing Business Process Modeling Notation (Object Management Group 2011) integrated with network-attack and mission-dependency graphs. Mission process models have also been integrated with information infrastructure, with validation and reasoning provided via ontology language (Barreto et al. 2012).

## 5.7 A Common Mission Modeling Language

A key research direction is to develop standard taxonomy for inferring mission dependencies, risks, and impacts (including resilience parameters) based on empirical studies. For example, this might include inferring recovery time of assets through system complexity indicators (subcomponent count, number of frameworks, users assigned to that component or process), asset value (cost) impact, data requirements for assets, or the presence of redundancy. This research direction dovetails with other forms of cybersecurity standardization (Martin 2009). Furthermore, the results of mission resilience analysis need to be effectively communicated to commanders, integrated through standard tactical dashboards, and mapped to geographic location as appropriate.

For assessing mission resilience to cyberattack and communicating analytic results to military commanders, a common mission modeling taxonomy and language are needed, which represent both mission vignettes and cyber entities. Standards such as Coalition Battle Management Language (C-BML) (NATO 2012) define information (orders, plans, reports, requests, etc.) that can be readily processed by command and control systems, simulation systems, or interfaces to automated forces. Serving as an interoperability standard based on Extensible Markup Language (SISO 2013), the focus of C-BML is to convey a commander’s intent. Other efforts (in both the US and UK) have focused more on developing a Cyber Mission Impact Assessment (CMIA) modeling standard for representing mission dependencies, risks, and impacts.

There are currently 2 CMIA standards that both allow the mission impact of potential cyberattacks on large military sociotechnical systems of systems to be modeled:

1) The US approach (Musman et al. 2013), developed by MITRE and based on BPMN, allows for explicit modeling of temporal resilience characteristics and information resources within a general-purpose approach that can compute measures of effectiveness at a mission level for specific categories of cyber impacts. This approach requires manual intervention to alter the mission model to reflect the cyber effect of the incident and repeated runs of the simulation to reflect the normal variations in mission instances.

2) The UK approach (Lang and Madahar 2017), developed by Dstl in collaboration with RJD Technology and based on the Unified Modeling Language (Object Mangement Group 2015), focuses on capturing networks of computer information system elements and defining mission device associations with critical mission components, termed operational technology (OT). Analysis scripts allow for the identification of highly connected (and therefore potentially critical) systems, potential attack paths between attack surfaces and critical systems, and the impact on business processes of a successful cyberattack. This approach was developed to support the analysis of individual military platforms and critical national infrastructure as part of cyber vulnerability investigations, and therefore requires an integration architecture to support analysis across multiple CMIA models to represent whole military deployments and cyber terrains. However, it does lend itself to the reuse of individual models to reduce the overall modeling burden.

The question of how to present the analysis results to mission operators has not been addressed in either the UK or US CMIA programs, and in both cases, the outputs of the analysis are highly dependent on the skills and knowledge of the CMIA analyst and the availability and accuracy of the technical information used to construct the models.

In the UK, the Joint User Cyber Mission Planning program (Waldock et al. 2017) is developing a concept demonstrator to combine some of the previously mentioned modeling concepts with advanced cyberattack and mission impact assessment (MIA) analysis algorithms, to allow tactical military personnel to plan and conduct missions involving cyber operations. In addition, the latest user interaction and visualization technologies are being trialed to effectively capture a commander's intent through a tactical map-based dashboard and communicate the results of analysis to military personnel at different levels.

Mission vignettes and computer networks (including software and vulnerabilities) are being modeled as a unified, scalable connected property-graph, allowing mission dependencies and resilience parameters to be explicitly modeled at varying levels of detail (LODs), as well as the application of MIA methods designed to assess cyber resilience (Dudman et al. 2017). UK CMIA model interoperability is being integrated to support the automatic generation of mission vignettes composing OT (critical mission components) for previously modeled military systems. The vignette is augmented with mission objectives and effects as military personnel interact with the variety of interfaces available to create and assess courses of action. Entire computer networks are imported separately from network information systems, so device associations are required to enable detailed automated cyber-resilience analysis. Joint military symbology (DOD 2014) has been extended to show the location of cyber entities (e.g., unknown entities, friendly defensive cyber sections, and hostile offensive cyber squads) in familiar NATO format.

Further research is needed for the reuse of existing system models and extensions to existing symbology. Questions about improved modeling of the human element of a cyberattack and deriving detailed temporal resilience parameters in a tactical situation also need to be addressed. Any common mission modeling language must include the required elements to capture the commander's intent, OT, mission dependencies, computer networks, resilience parameters, and the necessary analysis support, while minimizing information requirements in a tactical setting. The development of a unified CMIA modeling language could provide systems that utilize property-graph analysis techniques with detailed mission modeling system and effects templates at varying LODs, and also support advanced human-centric cyberattack and mission impact assessment techniques. Further research is also required into the LODs that would be required of such models to support the different levels of command.

## 5.8 Summary and Conclusions

In this section, we examined the role of mission models (e.g., mission-dependency graphs) for cyber resilience in military environments. Such models provide a framework that focuses resilience efforts on assuring missions and a mission-centric context for situational understanding in the face of cyberattacks. In terms of strategic directions for future research and development, it is important to consider automated methods for building such dependency models, to help reduce costs. Because resilience in cyberspace is inherently a time-dependent problem, mission models need to incorporate the dynamic nature of mission dependencies and network environments. Furthermore, the modeling abstractions and LODs need to be driven by mission requirements. Standardization efforts in the area of mission models for cyber resilience can also help in reducing costs and improving modeling accuracy and consistency.

# 6. Dynamic Defense in a Cyber-Resilient System

Authors: Nathaniel Evans, Luis Muñoz-González, Nandi Leslie, Donald W French, Donald Woodard, Kerry Krutilla, and Amanda Joyce

To understand the cyber resilience of a CPS, we must consider the system's operational responsibility, whether the engineered system can adapt to adversarial, heterogeneous environments and continue to perform as designed, how quickly the system can repair itself following a cyberattack, and what network security mechanisms exist that are adaptive and can serve as dynamic defenses for CPS entities. Uniform and static approaches are not useful for cyber resilience and risk assessments. Although the applications that run on a computer host or networked device operate as a series of straightforward steps, the interaction between these applications and hardware components can lead to complex performance dynamics (Mytkowicz et al. 2009). For example, even though identical code is run on Intel Pentium 4-based and Intel Core2-based computers, the Intel Pentium 4-based computer can exhibit periodic cache-miss behavior; while aperiodic, and potentially chaotic, behavior is observed for the Intel Core 2-based computer (Berry et al. 2006). Both of these processors have the same ISA specifications and manufacturer.

Cyber systems (e.g., intrusion prevention systems, intrusion detection systems [IDSs]) often are not scalable to cope with the size of networked devices, when considering all the components simultaneously in the system—this especially holds for lightweight, wearable devices in an Internet of Things (IoT) and large modern infrastructures of CPSs or industrial IoT. On the other hand, frequent changes in the system require the re-computation of the analyses from scratch, which is impractical and computationally expensive (or intractable) in many cases. There are a number of computational modeling techniques that can be useful to address the CPS security dynamics at multiple scales: compositional analysis tools and machine-learning algorithms can help to cope with the aforementioned challenges—in addition, developing and assessing measures of system-level reliability are essential for our understanding of the intensity of the cyberattack and how quickly (or whether) the damage can be remediated such that the CPS entity returns to some level of functioning—this process refers to cyber-resilience assessments. For example, Rodrigues et al. (2015) propose a compositional model capable of measuring reliability and modeling failure scenarios by composing models of its subcomponents. This allows efficient updates to the model, when new components are added or existing components are reconfigured or removed.

Furthermore, network performance for CPS entities can be analyzed at different levels of granularity, where only parts of the system are considered. This helps clarify the cyber risks of the system, specifically under the presence of a failure or cyberattack. Analyses at different scales can also be combined to produce aggregate measures of cyber risk and resilience, thus reducing the computational complexity of the analysis, enabling the parallelization of the computation of such measures and a more efficient updating of the analysis, when changes are observed in specific parts of the system. This kind of compositional analysis introduces a tradeoff among accuracy, level of granularity, and computational complexity. Since a gross modeling of the components of the system will reduce the computational complexity, the accuracy of the measurement of risk and resilience can be affected.

For CPSs, it is also important for our cyber-resilience assessments to measure and forecast key cyber-risk characteristics, such as the timing and type of exploit and the number of successful cyberattacks. In this context, there exists visualization tools capable of providing representations of the attack paths an attacker can follow to compromise valuable assets in the system at different levels of granularity (Noel and Jajodia 2004; Noel et al. 2005). This granularity allows the user to manage the complexity of graph representations, in turn helping humans to better understand the levels of cyber risk. Muñoz-González et al. (2017) showed that approximate measures of risk in Bayesian attack graphs can be as accurate as exact analysis techniques, while imposing a lower computational burden. However, the compositional generation and analysis of attack graphs remains an open problem.

Using attack graphs, the timing of lateral moves in a cyberattack can be projecting or forecasted—this type of predictive modeling is critical for cyber-resilience assessments (Yang et al. 2014). Attack graphs have also been valuable for computing the optimal cyber policy (Oldehoeft 1992) and control laws for implementation in cyber systems (e.g., IDSs, firewalls) for industrial control systems (ICSs)—security policies define the goals and elements of an organization's cyber systems (Strapp and Yang 2014; Yang et al. 2014). Furthermore, using Poisson and negative binomial generalized linear models, Leslie et al. (2017) show that network security policy violations greatly influence cyber vulnerabilities for an organization or company's network—specifically, the number of such violations is a strong predictor of the number of successful cyberattacks. In ICSs (e.g., power systems), resilient and robust controllers can be defined using discrete-time hybrid models that are based on value iteration and linear matrix inequalities, and these models have been used to study the effect of the IDSs' cybersecurity policies on the ICS under a denial-of-service attack (Yuan et al. 2013).

Another, more traditional way to test cyber resilience would be to engage a red team to measure mean time to recovery (MTTR) on a baseline of an infected system with no post-breach tools in use, then test again after introducing dynamic solutions. A cross reference MTTR with a cost-benefit analysis of adding these tools would be a valuable maturity measure.

Figure 8 outlines adaptive cyber-resilience mechanisms.

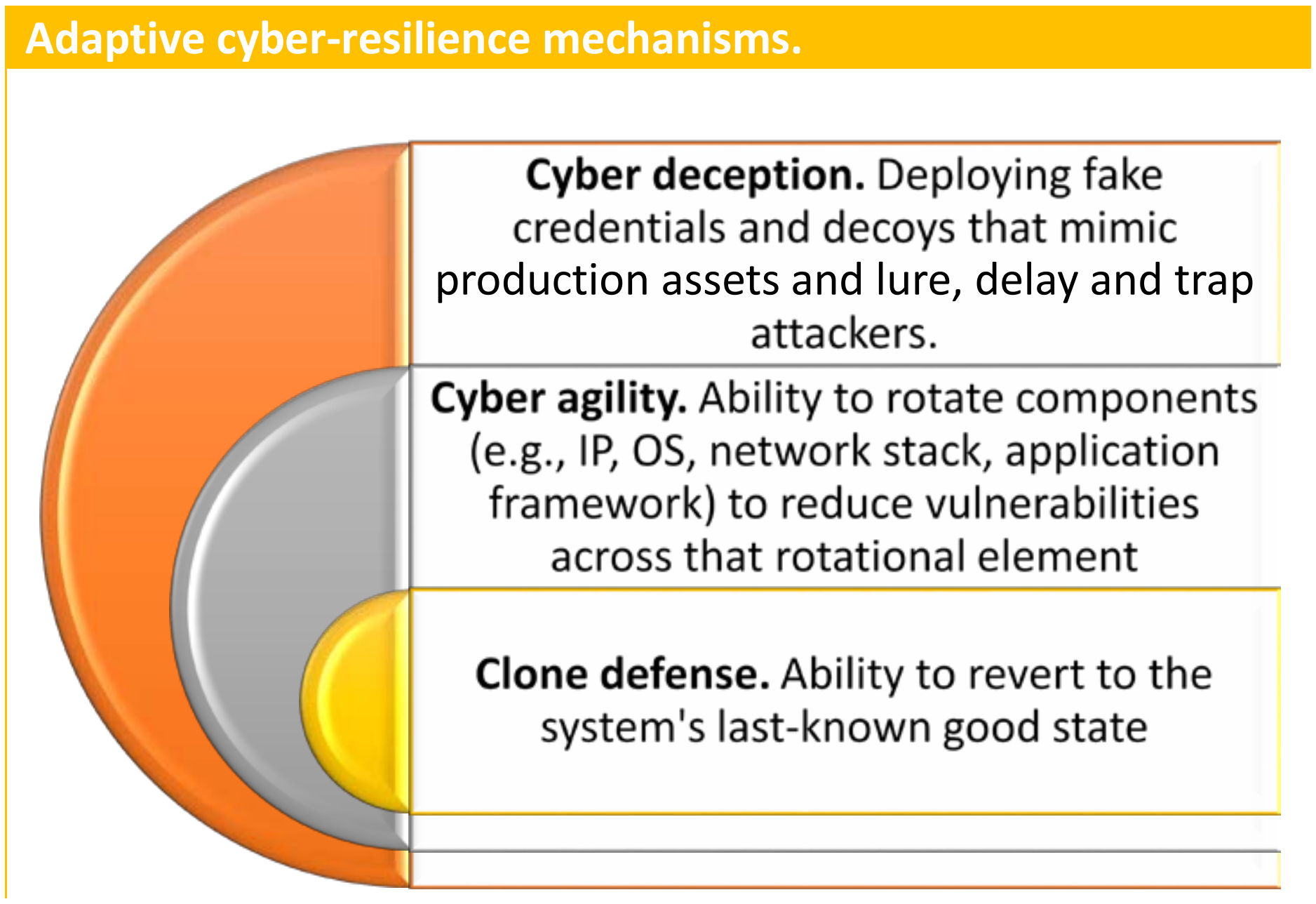

**Fig. 8 Adaptive cyber-resilience mechanisms**

Key cyber-resilience mechanisms are cyber deception, MTD or cyber agility, and clone defense—these processes provide dynamic CPS defense. *Cyber deception* works in networks, datacenters, cloud, supervisory control and data acquisition ICS systems, and the IoT. It is also capable of gathering forensics and reporting on the behavior of the attacker to understand its intent. It can easily scale as the dimensionality of the system grows. *MTD* is based on a simple premise that a moving target is harder to attack than a stationary target (Mellinger 2016). As a CPS gets more complex, the ability to implement a moving target solution gets more complicated to measure effectiveness. *Clone defense* is reverting to a point of a last known good state of the system. This can be accomplished from snapshots taken on a virtual machine or via a known good backup system. As a dynamic system (e.g., CPS, IoT) surges in size, the ability to revert to a known good state is demanding.

Cyberattacks are frequent and are often stealthy, requiring CPSs and the IoT to be resilient to their intended affects. Cyber-resilience approaches that are rules-based and static are insufficient for connected devices in dynamic, contested environments, where the cyber threats are difficult to predict. Instead, we need cyber-resilience assessments for CPSs and the IoT based on computational algorithms and models that predict and forecast cyber-risk factors, such as identifying the targeted device (or host computer system) of a cyberattack, timing of the attack, and frequency of intrusions by type. In addition, examining the impacts of dynamic, strategic mechanisms for CPS defense—these include cyber deception, MTD, and clone defenses—can improve cyber resilience by adapting to dynamic exploit-vulnerability pairings in adversarial environments.

## 7. Conclusion

Much work remains in understanding the fundamental properties of cyber resilience. Cyber resilience is an increasingly discussed but as yet not well-understood concept. One approach to enhancing and clarifying the concept is to identify connections between well-known concepts of information assurance and the resilience of an information system to attack. Well-formed properties must be time-independent, quantifiable, and calculated rather than measured. It is possible and useful to develop a mapping of categories of system properties to resilience objectives.

Approaches to measuring and modeling cyber resilience should not be confused with those developed for risk. While both risk and resilience methods aim to strengthen a system against potential threats, they have different foundations; therefore, quantification of risk and resilience should be approached differently. Unlike risk models, resilience models aim to determine whether and how quickly a system's critical functionalities can be restored given an unspecified event in order for the system to recover to a normal, functioning state. Cybersecurity networks are complex and interconnected networks in which known and unknown threats may impair the integrity of the system as a whole. If the structure of a network's interconnections and critical functionalities is well understood, resilience models help simulate how each state of the network will be impacted by an attack and how cascading effects will influence the resilience of the whole system. While existing risk and resilience modeling approaches are implemented, advanced quantification tools should be further explored to develop a comprehensive understanding of how resilience modeling can help enhance cyber resilience.

Modeling of a mission is critical to modeling and enhancing cyber resilience. Because of the challenges and costs of assessing and improving resilience across a military organization, such activities should be aligned to specific mission requirements. However, in the current state of practice, developing and continually maintaining mission models remains an expensive process. Such models should provide a framework that focuses resilience efforts on assuring missions and a mission-centric context for situational understanding in the face of cyberattacks. In terms of strategic directions for future research and development, it is important to consider automated methods for building such dependency models, to help reduce costs. Because resilience in cyberspace is inherently a time-dependent problem, mission models need to incorporate the dynamic nature of mission dependencies and network environments. Furthermore, the modeling abstractions and levels of detail need to be driven by mission requirements. Standardization efforts in the area of mission models for cyber resilience can also help in reducing costs and improving modeling accuracy and consistency.

Engineering-driven actions are necessary to develop more resilient systems by integrating cybersecurity/SSE to that of the well-known SE process. The Cyber Resiliency Engineering Framework developed by MITRE provides an overview of how to structure cyber-resiliency capabilities by addressing the goals, objectives, and practices in alignment with the "adversary activities" that occur within each capability to reflect the intent and potential actions that the capabilities are intended to protect. The mapping to the Cyber Resiliency Engineering Framework and the Start Secure, Stay Secure, and Return Secure concept are a useful starting point toward further development of a NATO cyber resilience-by-design capability.

A particularly important path toward cyber resilience is implementation of dynamic defense. Uniform and static approaches are not useful for cyber resilience and risk assessments. Key cyber-resilience mechanisms are cyber deception, MTD or cyber agility, and clone defense—these processes provide dynamic defense.

## 8. References

Alexeev A, Henshel DS, Levitt K, McDaniel P, Rivera B, Templeton S, Weisman M. Constructing a science of cyber-resilience for military systems. NATO IST-153 Workshop on Cyber Resilience; 2017 Oct 23–25; Munich, Germany.

Barreto A, Costa P, Yano E. A semantic approach to evaluate the impact of cyber actions to the physical domain. In 7th International Conference on Semantic Technologies for Intelligence, Defense, and Security; 2012 Oct 24–25; Fairfax, VA.

Berry H, Gracia Pérez D, Temam O. Chaos in computer performance. Chaos: An Interdisciplinary Journal of Nonlinear Science. 2006;16(1):013110.

Bodeau D, Graubart R. Cyber Resiliency Engineering Framework. McClean (VA): The MITRE Corporation; 2011 Sep. Report No.: MTR110237.

Bodeau DJ, Graubart RD. Cyber resiliency design principles: selective use throughout the lifecycle and in conjunction with related disciplines. McClean (VA): The MITRE Corporation; 2017 Jan [accessed 2018]. https://www.mitre.org/sites/default/files/publications/PR%2017-0103%20Cyber%20Resiliency%20Design%20Principles%20MTR17001.pdf.

Briguglio L, Cordina G, Farrugia N, Vella S. Economic vulnerability and resilience: concepts and measurements. Oxford Devel Studies. 2009;37(3):229–247. doi:10.1080/13600810903089893.

Cimellaro GP, Dueñas-Osorio L, Reinhorn AM. Introduction to special issue on resilience-based analysis and design of structures and infrastructure systems. J Structural Engineering. 2016;142(8).

Cyber Command System (CyCS). McLean (VA): The MITRE Corporation; 2017 May. [accessed 2018]. http://www.mitre.org/research/technology-transfer/technology-licensing/cyber-command-system-cycs.

Department of Homeland Security (US). DHS risk lexicon. Washington (DC): Department of Homeland Security (US), Risk Steering Committee; 2008 Sep [accessed 2018]. https://www.dhs.gov/xlibrary/assets/dhs_risk_lexicon.pdf.

[DOD] MIL-STD-2525D. Interface standard: joint military symbology. Washington (DC): Department of Defense; 2014.

Dudman T, Waldock A, Barrington S. JUMP: modelling and simulation of cyber resilience for mission impact assessment. NATO IST-153 Workshop on Cyber Resilience; 2017 Oct 23–25; Munich, Germany.

[HDA]. Intelligence preparation of the battlefield/battlespace. Washington (DC): Headquarters. Department of the Army; Quantico (VA): Marine Corps (US); 2014 Nov. Report No.: ATP 2-01.3/MCRP 2-3A.

Heinbockel W, Noel S, Curbo J. Mission dependency modeling for cyber situational awareness. In NATO IST-148 Workshop on Cyber Defence Situation Awareness; 2016 Oct 3–4; Sofia, Bulgaria.

Holling CS. Resilience and stability of ecological systems. Annual Rev Ecology and Systematics. 1973;4:1–23.

Kott A, Abdelzaher T. Resiliency and robustness of complex systems and networks. Adapt Dyn Resilient Syst. 2014;67:67–86.

Lang C, Madahar B. Understanding the mission impact of a cyber attack in a system of systems environment. In NATO IST-156 Workshop on Modelling and Simulation S&T: Critical Enabler for Cyber Defence; 2017 Jul 18–21; Portsmouth, UK.

Leslie NO, Harang RE, Knachel LP, Kott A. Statistical models for the number of successful cyber intrusions. J Defense Modeling and Sim. 2017;15(1):49–63. doi: 10.1177/1548512917715342.

Linkov I, Eisenberg DA, Plourde K, Seager TP, Allen J, Kott A. Resilience metrics for cyber systems. Environment Systems and Decisions. 2013;33(4):471–476.

Loscocco P, Smalley S. Integrating flexible support for security policies into the Linux operating system. Fort Meade (MD): National Security Agency; 2001. [accessed 2017 Oct 31]. https://www.nsa.gov/resources/everyone/digital-media-center/publications/research-papers/assets/files/flexible-support-for-security-policies-into-linux-feb2001-report.pdf.

Martin R. Making security measurable and manageable. CrossTalk: The Journal of Defense Software Engineering. 2009 Sep/Oct.

Mellinger A. SEI insights. Pittsburgh (PA): Carnegie Mellon University; 2016 Apr [accessed 2018]. https://insights.sei.cmu.edu/sei_blog/2016/04/a-platform-for-dynamic-defense-technologies.html.

Muñoz-González L, Sgandurra D, Paudice A, Lupu EC. Efficient attack graph analysis through approximate inference. ACM Transactions on Privacy and Security (TOPS). 2017;20(3).

Musman S, Agbolosu-Amison S. A measurable definition of resiliency using “mission risk” as a metric. McClean (VA): The MITRE Corporation; 2014. Report No.: MTR140047.

Musman S, Temin A, Tanner M, Fox R, Pridemore B. Evaluating the impact of cyber attacks on missions. M&S Journal. 2013 Summer;8(2):25–35.

Musman S. Automagical dependency mapping. McClean (VA): The MITRE Corporation; 2017.

Mytkowicz T, Diwan A, Bradley E. Computer systems are dynamical systems. Chaos: An Interdisciplinary Journal of Nonlinear Science. 2009;19(3):033124.

Nejib P, Yakabovicz E. NATO resilience by design: enhancing resilience through cyber systems engineering. NATO IST-153 Workshop and Report; 2017 Oct 23–25; Munich, Germany.

Noel S, Bodeau D, McQuaid R. Big-data graph knowledge bases for cyber resilience. NATO IST-153 Workshop on Cyber Resilience; 2017 Oct 23–25; Munich, Germany.

Noel S, Harley E, Tam KH, Gyor G. Big-data architecture for cyber attack graphs: representing security relationships in NoSQL graph databases. In IEEE Symposium on Technologies for Homeland Security (HST); 2015; Boston, MA.

Noel S, Harley E, Tam KH, Limiero M, Share M. CyGraph: graph-based analytics and visualization for cybersecurity. In Cognitive computing: theory and applications, handbook of statistics 35. Amsterdam (The Netherlands): Elsevier; 2016.

Noel S, Jacobs M, Kalapa P, Jajodia S. Multiple coordinated views for network attack graphs. In Visualization for Computer Security. (VizSEC 05). IEEE Workshop; 2005 Oct 26; Minneapolis, MN. p. 99–106.

Noel S, Jajodia S. Managing attack graph complexity through visual hierarchical aggregation. In Proceedings of the 2004 ACM Workshop on Visualization and data mining for computer security; 2004 Oct. p. 109–118.

Noel S, Jajodia S. Metrics suite for network attack graph analytics. In 9th Annual Cyber and Information Security Research Conference (CISRC); 2014; Oak Ridge National Laboratory, Oak Ridge, TN.

Noel S, Ludwig J, Jain P, Johnson D, Thomas RK, McFarland J, King B, Webster S, Tello B. Analyzing mission impacts of cyber actions (AMICA). In NATO IST-128 Workshop on Cyber Attack Detection, Forensics and Attribution for Assessment of Mission Impact; 2015; Istanbul, Turkey.

North Atlantic Treaty Organisation (NATO) Research and Technology Organisation (RTO). Coalition Battle Management Language (C-BML). Brussels (Belgium): NATO; 2012. Report No.: TR-MSG-048.

Object Management Group. Business Process Model and Notation (BPMN). Needham (MA): Object Management Group; 2011.

Object Mangement Group. OMG Unified Modeling Language (OMG UML). Needham (MA): Object Management Group; 2015 Mar.

Oldehoeft A. Foundations of a security policy for use of the national research and educational network. Gaithersburg (MD): Computer Security Resource Center, National Institute of Standards and Technology; 1992 Feb. [accessed 2018]. https://csrc.nist.gov/publications/detail/nistir/4734/final.

Rodrigues P, Lupu E, Kramer J. Compositional reliability analysis for probabilistic component automata. In Proceedings of the 7th International Workshop on Modeling in Software Engineering. IEEE Press; 2015 May. p. 19–24.

Schulz A, O’Gwynn D, Kepner J, Trepagnier P. Dynamically correlating network terrain to organizational missions. NATO IST-153 Workshop on Cyber Resilience; 2017 Oct 23–25; Munich, Germany.

Simulation Interoperability Standards Organisation (SISO). Standard for Coalition Battle Management Language (C-BML) phase 1. Atlanta (GA): SISO; 2013.

Strapp S, Yang SJ. Segmenting large-scale cyber attacks for online behavior model generation. In Proceedings of 2014 International Conference on Social Computing, Behavioral-Cultural Modeling & Prediction (SBP14); 2014 Apr 1–4; Washington, DC. p. 169–177.

Trepagnier P, Schulz A. Mission assurance as a function of scale. NATO IST-128 Workshop on Cyber Attack Detection, Forensics and Attribution for Assessment of Mission Impact; 2015 June 15–17; Istanbul, Turkey.

Trimintzios P. Measurement frameworks and metrics for resilient networks and services. Heraklion (Greece): European Network and Information Security Agency; 2011 Feb 1 [accessed 2018]. https://www.enisa.europa.eu/publications/metrics-tech-report/at_download/fullReport. Discussion draft.

Waldock A, Dudman T, Harold SJ, Barrington S. JUMP: concept demonstrator for cyber mission planning. NATO IST-156 Workshop on Modelling and Simulation S&T: Critical Enabler for Cyber Defence; 2017 Jul 18–21; Portmouth, UK.

Watson J-P, Guttromson R, Silva-Monroy C, Jeffers R, Jones K, Ellison J, Rath C, Gearhart J, Jones D, Corbet T, Hanley C, Walker LT. Conceptual framework for developing resilience metrics for the electricity, oil, and gas sectors in the United States. Albuquerque (NM): Sandia National Laboratories; 2015 Sep [accessed 2018]. Report No.: SAND2014-18019. https://cfwebprod.sandia.gov/cfdocs/CompResearch/docs/EnergyResilienceReportSAND2014-18019o.pdf.

Yang SJ, Du H, Holsopple J, Sudit M. Attack projection. In Cyber defense and situational awareness. Berlin (Germany): Springer International Publishing; 2014. p. 239–261.

Yuan Y, Zhu Q, Sun F, Wang Q, Başar T. Resilient control of cyber-physical systems against denial-of-service attacks. In Resilient Control Systems (ISRCS). 2013 6th International Symposium on IEEE; 2013 Aug 13–15; San Francisco, CA. p. 54–59.

## List of Symbols, Abbreviations, and Acronyms

| | |
|---|---|
| C-BML | Coalition Battle Management Language |
| CMIA | Cyber Mission Impact Assessment |
| CPS | cyber-physical systems |
| HTTP | Hypertext Transfer Protocol |
| ICS | industrial control system |
| ICT | information and communication technology |
| IDS | intrusion detection system |
| *imp* | impact of the attack measured by the decrease of performance |
| IoT | Internet of Things |
| LODs | levels of detail |
| MIA | mission impact assessment |
| MTD | moving target defense |
| MTTR | mean time to recovery |
| NATO | North Atlantic Treaty Organization |
| OT | operational technology |
| SE | systems engineering |
| SSE | systems security engineering |
| $t_{fp}$ | time to reach the final performance |
| $t_{ip}$ | time to restore the initial performance |
| $\Delta p$ | difference between the final and the initial performance |
| 1 (PDF) | DEFENSE TECHNICAL INFORMATION CTR DTIC OCA |
| 2 (PDF) | DIR ARL IMAL HRA RECORDS MGMT RDRL DCL TECH LIB |
| 1 (PDF) | GOVT PRINTG OFC A MALHOTRA |
| 1 (PDF) | ARL RDRL D A KOTT |